\documentclass[AMA,STIX1COL]{WileyNJD-v2}
\articletype{Research Article}%

\received{28 November 2018}
\revised{11 June 2019}
\accepted{25 June 2019}

\usepackage{lscape,rotating}
\usepackage{float,epsfig}
\usepackage{subfigure}
\usepackage{graphicx}

\newcommand\blfootnote[1]{%
  \begingroup
  \renewcommand\thefootnote{}\footnote{#1}%
  \addtocounter{footnote}{-1}%
  \endgroup
}

\begin{document}

\title{Sample size calculation for the Andersen-Gill model comparing rates of recurrent events}

\author{Yongqiang Tang, Ronan Fitzpatrick}

\address{Tesaro, 1000 Winter Street, Waltham, MA 02451, USA\newline
Statistical Solutions Ltd., 4500 Avenue
4000, Cork Airport Business Park, Cork,
T12 NX7D, Ireland}

\corres{Tesaro, 1000 Winter St, Waltham, MA 02451 \newline
E-mail: yongqiang\_tang@yahoo.com}

\abstract[Summary]{
Recurrent events arise frequently in biomedical research, where the subject may experience the same type of events more than once. 
 The Andersen-Gill (AG) model  has become increasingly popular in the analysis of recurrent events particularly when the event rate is not constant over time.
We propose a procedure for calculating the power and sample size for the robust Wald test from the AG model in superiority, noninferiority and equivalence clinical trials. Its performance is demonstrated by numerical examples.
Sample SAS code  is provided in the supplementary material.
}


\maketitle

\section{Introduction}
Recurrent events are frequently encountered in biomedical research, where the subject may experience the same type of events more than once. 
Examples include attacks in hereditary angioedema,  exacerbations in chronic obstructive pulmonary disease, bleeds in hemophilia,  relapses in multiple sclerosis,  and infections in 
chronic granulomatous disease (CGD). In clinical trials, the  recurrent events are commonly analyzed by the negative binomial (NB) regression  
\cite{wang:2009, aban:2008, tang:2015}. 
The NB regression assumes constant event rates over time, which may fail to hold in some applications \cite{matsui:2005,nicholas:2011,inusah:2010}.
The Andersen-Gill (AG) model \cite{ andersen:1982} provides a popular alternative tool for the analysis of recurrent events, and it allows arbitrary event rate functions.
The AG model  often yields similar treatment effect estimates (i.e. ratio of event rates between groups) to the NB regression in empirical studies
when the event rate is roughly constant over time  \cite{wang:2009}.

\blfootnote{The paper was published in Statistics in Medicine 2019 (Volume 38, Issue 24, Pages 4819 - 4827).
There was an error in Equation A4 in the appendix. It does not affect design 1, but appears to slightly overestimate the sample size for design 2 with staggered entry.
The result becomes better after the correction in the sense that 
the nominal power generally becomes closer to the simulated power for design 2.  The corrected contents were highlighted in red.}

Sample size calculation  is critical in  designing a clinical trial to ensure  sufficient power to detect an important treatment effect.
Sample size    methodology  has been well developed for the NB regression; Please see Tang  \cite{tang:2017aaa} and references therein.
Matsui \cite{matsui:2005} and  Song {\it et al} \cite{song:2008} derive sample size formulae 
for the robust log-rank test \cite{lawless:1995}, which is a nonparametric test suitable only for superiority trials. 
 In this paper, we propose a  power and sample size calculation procedure for the robust Wald test from the AG model  \cite{lin:2000}. It is applicable to superiority, noninferiority (NI) and equivalence trials.
Two designs are considered. In one design, the planned
treatment duration is the same for all subjects. In the other design, subjects are enrolled at different
calendar time, but administratively censored at the same calendar time.
We introduce the sample size procedure  in Section $2$, and assess its performance numerically in Section \ref{simulation}.

 \section{Power and sample size formulae}\label{method}
Andersen and Gill \cite{andersen:1982} provides a simple extension of the Cox proportional hazards model to the analysis of recurrent events.
Suppose $n$ subjects are randomized to either the active ($x_i=1$) or  control  ($x_i=0$) treatment in a clinical trial. Let $T_i$ be the follow-up time for subject $i$, 
$Y_i(t)=I(T_i\geq t)$   the indicator function that subject $i$ is still under observation at
time $t$,  and $N_i(t)$  the number of events experienced by subject $i$ by time $t$.  
 Inference for the event rate ratio $\exp(\beta)$ between treatment groups is based on the following partial likelihood
\begin{equation}\label{likelihood}
PL(\beta) = \prod_{i=1}^n \prod_{\{t:Y_i(t)=1\}} \left[\frac{ \exp(\beta x_i)}{\sum_{j=1}^n Y_j(t)  \exp(\beta x_j)}    \right]^{dN_i(t)}.
\end{equation}
An attractive feature of the AG model is that the baseline event rate function can be of arbitrary shape.  
 We assume a constant event rate ratio over time, but the AG model can handle time-varying treatment effects.

 To obtain the maximum likelihood estimate (MLE) $\hat\beta$, we  solve the score function
$$ U(\beta) =\frac{\partial \log[PL(\beta)  ]}{\beta} =\sum_{i=1}^n  \int_{0}^\tau [x_i -\bar{x}(\beta, t)] dN_i(t)=0,$$
where 
$S^{(k)}(\beta,t)= n^{-1} \sum_{i=1}^n Y_i(t) x_i^k \exp(\beta x_i)$,  $\bar{x}(\beta, t)= \frac{S^{(1)}(\beta,t)}{S^{(0)}(\beta,t)}$,
and $\tau$ is the maximum treatment duration in the trial.

If all covariates are time invariant (the covariates measured after randomization are rarely used to assess the treatment effect in clinical trials since the covariates may be affected by the treatment), 
the AG model  assumes that  the time increments between events are independent according to a Poisson process,
but the recurrent events are generally dependent within a subject \cite{wang:2009}. 
The Poisson-type assumption can be relaxed by using the sandwich variance estimator, and  the validity of this robust approach is justified by Lin {\it et al}  \cite{lin:2000} 
for arbitrary dependence structures among recurrent events if the proportional rate or mean assumption is met. For this reason, the robust approach is also called the proportional  rates/means model.
The sandwich variance estimate \cite{lin:2000} for $\hat\beta$ is $n^{-1}\hat{V}_\beta=n^{-1}\hat I_\beta^{-1} \hat \Sigma_\beta \hat I_\beta^{-1}$,
 where  $\hat\Lambda_0(t) =\sum_{i=1}^{n}\int_0^\tau [n S^{(0)}(\hat\beta,t)]^{-1}dN_i(t)$,
   $d\hat M_i(t)=dN_i(t)-Y_i(t) \exp(\hat\beta x_i) d\hat\Lambda_0(t)$, $\hat{U}_i =   \int_{0}^\tau [x_i -\bar{x}(\hat\beta,t)] d\hat{M}_i(t)$,
   $\hat \Sigma_{\beta} = n^{-1} \sum_{i=1}^{n}\hat{U}_{i}^2$,  and $\hat I_\beta = n^{-1} \sum_{i=1}^{n} \int_0^\tau [x_i-\bar{x}(\hat\beta,t)]^2 Y_i(t) \exp(\beta x_i)d\hat\Lambda_{0}(t)
= n^{-1} \sum_{i=1}^{n} \int_0^\tau [\bar{x}(\hat\beta,t)-\bar{x}^2(\hat\beta,t)]dN_{i}(t)\,$.
 The  two-sided  $100(1-\alpha)\%$ confidence interval (CI) for $\beta$ is 
$$[c_l,c_u]=[\hat\beta-z_{1-\alpha/2} \sqrt{n^{-1}\hat{V}_\beta}, \hat\beta+z_{1-\alpha/2} \sqrt{n^{-1}\hat{V}_\beta}],$$
where $z_p$ is the $p$-th percentile of the standard normal distribution $N(0,1)$.

In the sample size determination,  we assume  a mixed Poisson process (MPP) model  \cite{matsui:2005, song:2008,tang:2017c} for the event process.
Let $\Lambda_g(t) =\text{E}[N_i(t)|x_i=g]$ be the mean event function for group $g$.  The MPP introduces a random effect $\epsilon_i$ with mean $1$ and variance $\kappa_g$ for each subject.  
Given $\epsilon_i$, the subject in group $g$ follows a Poisson process with mean function $\epsilon_i \Lambda_g(t)$.
Subjects with $\epsilon_i>1$ ($\epsilon_i <1$) tend to experience more (less) events than the average in the population.
The dispersion parameter $\kappa_g$ measures the between-subject heterogeneity.
Inclusion of important risk factors in the model may reduce heterogeneity \cite{tang:2017c}. The MPP
provides a natural way to handle overdispersion in recurrent events in  that the variance of $N_i(t)$ is larger than its mean \cite{tang:2017c}.
The mixing distribution for the random effect $\epsilon_i$ is unspecified  in the AG model. 
The NB regression uses a gamma  mixing distribution, and the event count  $N_i(t)$ follows the NB distribution \cite{tang:2015}.

In Appendix \ref{proofsig}, we show that  $\hat{V}_\beta$ converges in probability to $V_\beta$
\begin{equation}\label{sigma1}
V_\beta = \frac{ p_1[A_1+\kappa_1 B_1]+p_0[A_0+\kappa_0 B_0]}{\left( \int_{0}^\tau  \frac{ [p_0\pi_0(t) ]\,[p_1\pi_{1}(t) \exp(\beta)] }{p_0\pi_{0}(t) +p_1\pi_{1}(t) \exp(\beta)}d\Lambda_0 \right)^2},
\end{equation}
where $p_g$ is the proportion of subjects randomized to treatment group $g$,
$\pi_g(t)$ is the probability that a subject in group $g$ remains in the study at time $t$, 
$\omega_{0}(t) = \frac{p_1\pi_{1}(t) \exp(\beta) }{p_1\pi_{1}(t) \exp(\beta)+p_0\pi_{0}(t)}$, $\omega_1(t)=1-\omega_{0}(t)$,
$A_g = \int_{t=0}^\tau \omega_{g}^2(t) \pi_{g}(t)d\Lambda_{g}(t)$, and
$B_g =2 \int_{t=0}^\tau\left[\int_{s=0}^t \omega_{g}(s) d\Lambda_g(s)\right] \pi_g(t) \omega_{g}(t)d\Lambda_g(t)$. We allow the loss to follow-up distribution
$G_g(t)=1-\pi_g(t)$ and the dispersion parameter $\kappa_g$ to differ between the two treatment groups.

At the design stage, it is often reasonable to assume the same dropout distribution in the two treatment groups (i.e. $\pi_1(t)=\pi_0(t)$ for all $t$), and $V_\beta$ reduces to
\begin{equation}\label{sigma2}
V_\beta = \frac{1}{p_1E_1}+\frac{1}{p_0E_0} +2 \left(\frac{\kappa_1}{p_1}\frac{F_1}{E_1^2}+\frac{\kappa_0}{p_0} \frac{F_0}{E_0^2}\right)= \left[\frac{1}{p_1\exp(\beta)}+\frac{1}{p_0}\right]\frac{1}{E_0} + \left[\frac{\kappa_1}{p_1}+ \frac{\kappa_0}{p_0}\right] \frac{2F_0}{E_0^2},
\end{equation}
where $E_g =\int \pi_g(t)d\Lambda_g(t)$ and  $F_g  = \int_{t=0}^\tau \pi_g(t)\Lambda_g(t)d\Lambda_g(t)$.  In general, formula \eqref{sigma1} can be well approximated by the term
between the two equal signs in formula \eqref{sigma2} if the dropout distribution differs between the two groups.

In Appendix \ref{analytvar}, we provide analytic expressions of $E_g$ and $F_g$ for the Weibull and piecewise constant event rate functions 
when the dropout pattern is identical in the two groups  in two types of clinical trial designs. In  practical applications,
almost any event rate function can be approximated reasonably well by the piecewise constant  function.

\subsection{Superiority and NI  trials}
Suppose a lower event rate is desirable. In both superiority and NI trials, the hypothesis can be written as
\begin{equation}\label{testsupni}
 H_0: \exp(\beta) \geq M_0  \text{ or }  \beta\geq \log(M_0)  \text{ \it  versus  }  H_1:  \exp(\beta) < M_0 \text{ or }  \beta< \log(M_0).  
\end{equation}
In a superiority trial, the objective is to demonstrate that the experimental treatment can lower the event rate, and we set $M_0=1$. 
The NI trial aims to  show that the experimental treatment is not worse than the standard control treatment by $M_0$, where $M_0>1$ is  the prespecifed NI margin on the rate ratio.

The power for test \eqref{testsupni} is given by
\begin{eqnarray}\label{powerni}
\begin{aligned}
\Pr(c_u<&\log(M_0)) 
=\Pr\left[ Z < \frac{-z_{1-\alpha/2}\sqrt{n^{-1}\hat{V}_\beta }-\beta+\log(M_0) }{\sqrt{n^{-1} V_\beta}} \right]
\approx \Phi\left[\frac{\sqrt{n}|\log(M_0)-\beta|}{\sqrt{ V_\beta}}-z_{1-\alpha/2}\right],
\end{aligned}
\end{eqnarray}
where $Z=(\hat\beta-\beta)/\sqrt{n^{-1} V_\beta}$ is asymptotically distributed as $N(0,1)$. 
The required sample size  is
\begin{equation}\label{sizeni}
n= \frac{(z_{1-\alpha/2}+z_P)^2 V_\beta}{[\log(M_0)-\beta]^2}.
\end{equation}

As mentioned in Tang \cite{tang:2015},
 Equation \eqref{sizeni}
is identical to the upper size bound of Tang \cite{tang:2015, tang:2018} for the NB regression (the dispersion parameter may differ between the two groups in Tang \cite{tang:2018})
 under the assumption of constant event rates if the dropout pattern is the same in the two groups
since $F_0=\lambda_0^2 \text{E}(T_{i}^2)/2$, $E_0=\lambda_0 \text{E}(T_{i})$, and
$$ V_\beta =  \left[\frac{1}{p_1\exp(\beta)}+\frac{1}{p_0}\right]\frac{1}{\lambda_0 \text{E}(T_{i})} + \left[\frac{\kappa_1}{p_1}+ \frac{\kappa_0}{p_0}\right] \frac{\text{E}(T_{i}^2)}{\text{E}^2(T_{i})}.$$
In this special situation, the AG model is almost  as powerful as the NB regression when the  variation in the patients' follow-up time $T_i$ is small, and
the two models yield the same power if all subjects have the same follow-up time $T_1=\ldots=T_n$.
 However, the AG model does not require specifying the mixing distribution.

The NI test is one-sided, and the actual type I error is $\alpha/2$. 
In  superiority trials, a two-sided test (i.e $H_0: \exp(\beta)=1$ {\it vs} $H_1: \exp(\beta)\neq 1$) is often used in practice.
Formulae \eqref{powerni} and \eqref{sizeni}  can be used for the two-sided test since  
there is little chance that the observed outcomes will be significantly better in the 
control group than in the experimental group if the experimental treatment is truly more effective than the control treatment \cite{tang:2018ancova}.
The power and sample size formulae \eqref{powerni} and \eqref{sizeni} remain the same  if higher event rates indicate better health ($M_0\leq 1$) and 
the experimental treatment is truly superior or clinically noninferior to the control treatment in improving the event rate.

\subsection{Equivalence trials}
In an equivalence trial, the objective is to demonstrate that the experimental treatment is neither superior nor inferior to the standard control treatment.
If the   $100(1-\alpha)\%$  CI for $\exp(\beta)$  lies completely  within the interval $[M_l,M_u]$, 
we can claim   clinical equivalence of the two treatments, where $M_l<1$ and $M_u>1$ are the prespecified margins.
 The hypothesis is
$$ H_0: \exp(\beta) \geq M_u \text{ or } \exp(\beta) \leq M_l   \text{ \it versus }  H_1: M_l < \exp(\beta)< M_u. $$
The equivalence test can be viewed as the two one-sided tests and the type I error is $\alpha/2$. The power is given by
\begin{eqnarray}\label{powerequiv}
\begin{aligned}
P &=  \Pr( \hat\beta+z_{1-\alpha/2} \sqrt{n^{-1}\hat{V}_\beta}<\log(M_u)  \text{ and }  \hat\beta-z_{1-\alpha/2} \sqrt{n^{-1}\hat{V}_\beta}>\log(M_l) ) \\
& \approx \Phi\left(\frac{\sqrt{n}[\log(M_u)-\beta]}{\sqrt{  V_\beta}}-z_{1-\alpha/2}\right) -
         \Phi\left(\frac{\sqrt{n}[\log(M_l)-\beta]}{\sqrt{ V_\beta}}+z_{1-\alpha/2}\right).
\end{aligned}
\end{eqnarray}
 Formula \eqref{powerequiv} assumes that  $z_{1-\alpha/2} \sqrt{n^{-1}\hat{V}_\beta}+ \log(M_l) <\log(M_u) - z_{1-\alpha/2} \sqrt{n^{-1}\hat{V}_\beta}$
 and hence $2z_{1-\alpha/2} \sqrt{n^{-1}\hat{V}_\beta} < \log(M_u/M_l)$, which may not hold with a positive probability.   Formula \eqref{powerequiv} 
 works well in large samples or when the estimated power is large, but generally underestimates the power  in small samples.
 The argument is the same as that for continuous outcomes \cite{tang:2018t,tang:2018ancova}.

The required sample size can be obtained by numerical inversion of the power formula \eqref{powerequiv}. In the special case when 
$\Delta =\log(M_u)-\beta=  \beta-\log(M_l) =\log(M_u/M_l)/2$, the sample size is given by
\begin{equation}\label{sizeequivbound}
n= \frac{(z_{1-\alpha/2}+z_{(1+P)/2})^2 V_\beta}{\Delta^2}.
\end{equation}

\section{Numerical examples}\label{simulation}
\subsection{Example 1}
We illustrate the sample size calculation for superiority trials by the analysis of a  CGD trial  \cite{lin:2000, matsui:2005}.
CGD is a rare immune system disorder characterized by recurrent pyogenic infections.
 A total of $128$ patients were randomized to  gamma interferon or placebo. 
The trial was terminated early for efficacy
on basis of an interim analysis of the time to the first infection.
 In the trial,  $14$ ($22.2\%$) out of $63$ treated patients  and $30$ ($46.2\%$) out of $65$ patients on placebo had at least one  infection.
Furthermore, $9$ placebo patients and $4$ treated patients experienced at least $2$ infections.

One  objective is to estimate the infection rate ratio between the two treatments. The NB regression
gives an estimate of $0.3566$ ($95\%$ CI: $[0.1934, 0.6575]$) while the AG model yields an estimate of $0.3338$ ($95\%$ CI: $[ 0.1814, 0.6143]$).
As evidenced by the exploratory analysis of  Matsui \cite{matsui:2005}, the rate of infections may not be constant over time. For this reason,
 the AG model is more appropriate for analyzing the CGD trial since it allows arbitrary event rate function.

Suppose we want to  design a new trial to assess the effect of a new experimental product on the infection rate.
We assume the event rate function is of Weibull form $\lambda_0(t)=\psi \nu t^{\nu-1}$ in the placebo arm, and the event rate ratio between the two treatments is constant $\lambda_1(t)/\lambda_0(t)=\exp(\beta)=0.6$ over time. 
We get the MLE  $(\hat\psi,\hat\nu,\hat\kappa)=(1.097^{1.221}, 1.221, 0.871)$  by fitting a NB process  \cite{tang:2017c} model to the data using the SAS NLMIXED  procedure on basis of
 the likelihood function given in Equation (20) of Dean and Balshaw \cite{dean:1997}.
Matsui \cite{matsui:2005} obtained similar point estimates based on the generalized estimating
equations (GEE) for the MPP \cite{dean:1997}.

To determine the sample size, we assume a common dispersion parameter and identical dropout pattern in the two groups.
We set $\psi=1.1$, $\nu=1.2$, $\kappa=0.8$, which are close to the MLE. The treatment allocation ratio is $p_1:p_0=1:1$ or $2:1$.
We also perform sensitivity analyses to calculate the sample sizes at alternative parameter values
$\kappa=0.4, 1.2$, $\psi=1.5$, $\nu=0.9$. Both design $1$ (planned treatment duration $\tau_c=1$ year for all patients) and design $2$ (accrual period $\tau_a=0.5$ year, additional treatment duration $\tau_c=1$ year,  constant enrollment rate $\eta=0$) are considered (please refer to Appendix \ref{analytvar} for  details).
 In both designs, the loss to follow-up distribution is exponential with mean $1/\delta=4$ years. 
 
Table \ref{supres} reports the sample size and power estimates at the target $90\%$ power and one-sided type I error $\alpha/2=0.025$.  
The empirical  power is evaluated based on $40,000$ trials. The data are simulated using Algorithm $2$ of Tang \cite{tang:2017c} and analyzed using the SAS PHREG procedure. 
 There is more than $95\%$ chance
that the simulated power lies within $2 \sqrt{ 0.9*0.1/40000}=0.3\%$ of the true power.
{\color{red} In both designs, \sout{In design $1$,} the simulated power is within $1\%$ of  the nominal power in nearly all cases.
\sout{The performance slightly deteriorates in design $2$ possibly because of larger variation in the follow-up time and higher overall dropout rate.}}

  \begin{table}[h]
\begin{center}
\begin{tabular}{l@{\extracolsep{5pt}}c@{\extracolsep{5pt}}c@{\extracolsep{5pt}}c@{\extracolsep{5pt}}c@{\extracolsep{5pt}}c@{\extracolsep{5pt}}c@{\extracolsep{5pt}}c@{\extracolsep{5pt}}c@{\extracolsep{5pt}}c@{\extracolsep{5pt}}c@{\extracolsep{5pt}}c@{\extracolsep{5pt}}c@{\extracolsep{5pt}}c@{\extracolsep{5pt}}cccc} \\\hline 
&&&  \multicolumn{6}{c}{Design $1$} &  \multicolumn{6}{c}{Design $2$}  \\ \cline{4-9}\cline{10-15}
&&&  \multicolumn{3}{c}{balanced size  }  & \multicolumn{3}{c}{unbalanced size}  & \multicolumn{3}{c}{balanced size } &  \multicolumn{3}{c}{unbalanced size } \\\cline{4-6}\cline{7-9}\cline{10-12}\cline{13-15}
  &&&total &\multicolumn{2}{c}{power ($\%$)}    & total  & \multicolumn{2}{c}{power ($\%$)} & total & \multicolumn{2}{c}{power ($\%$)}  & total  & \multicolumn{2}{c}{power ($\%$)}  \\\cline{5-6}\cline{8-9}\cline{11-12}\cline{14-15}
$\kappa$ & $\psi$ & $\nu$ & size  & nominal & SIM   & size &  nominal & SIM & size  & nominal & SIM   &size  & nominal & SIM \\\hline

      $0.4$&$1.1$&$0.9$&$ 289$&$90.05$&$91.03$&$ 304$&$90.00$&$89.05$&${\color{red} 256}$&${\color{red}90.01}$&${\color{red}90.67}$&${\color{red} 271}$&${\color{red}90.04}$&${\color{red}89.38}$\\
                              $ $&$ $&$1.2$&$ 294$&$90.01$&$90.68$&$ 310$&$90.03$&$89.33$&${\color{red} 251}$&${\color{red}90.10}$&${\color{red}90.75}$&${\color{red} 265}$&${\color{red}90.02}$&${\color{red}89.34}$\\
                              $ $&$1.5$&$0.9$&$ 231$&$90.12$&$90.78$&$ 244$&$90.03$&$89.13$&${\color{red} 207}$&${\color{red}90.05}$&${\color{red}90.63}$&${\color{red} 220}$&${\color{red}90.03}$&${\color{red}89.11}$\\
                              $ $&$ $&$1.2$&$ 235$&$90.07$&$90.45$&$ 249$&$90.07$&$88.90$&${\color{red} 204}$&${\color{red}90.13}$&${\color{red}90.54}$&${\color{red} 217}$&${\color{red}90.09}$&${\color{red}89.30}$\\
                              $0.8$&$1.1$&$0.9$&$ 358$&$90.02$&$90.77$&$ 382$&$90.01$&$89.15$&${\color{red} 328}$&${\color{red}90.08}$&${\color{red}90.67}$&${\color{red} 351}$&${\color{red}90.03}$&${\color{red}89.35}$\\
                              $ $&$ $&$1.2$&$ 365$&$90.03$&$90.77$&$ 390$&$90.05$&$89.23$&${\color{red} 324}$&${\color{red}90.03}$&${\color{red}90.48}$&${\color{red} 348}$&${\color{red}90.03}$&${\color{red}89.34}$\\
                              $ $&$1.5$&$0.9$&$ 300$&$90.07$&$90.39$&$ 322$&$90.03$&$89.05$&${\color{red} 278}$&${\color{red}90.02}$&${\color{red}90.52}$&${\color{red} 300}$&${\color{red}90.02}$&${\color{red}89.59}$\\
                              $ $&$ $&$1.2$&$ 306$&$90.08$&$90.74$&$ 328$&$90.01$&$89.34$&${\color{red} 277}$&${\color{red}90.04}$&${\color{red}90.60}$&${\color{red} 300}$&${\color{red}90.09}$&${\color{red}89.42}$\\
                              $1.2$&$1.1$&$0.9$&$ 428$&$90.06$&$90.74$&$ 460$&$90.01$&$89.49$&${\color{red} 399}$&${\color{red}90.05}$&${\color{red}90.38}$&${\color{red} 431}$&${\color{red}90.02}$&${\color{red}88.97}$\\
                              $ $&$ $&$1.2$&$ 436$&$90.04$&$90.59$&$ 469$&$90.01$&$89.55$&${\color{red} 398}$&${\color{red}90.06}$&${\color{red}90.24}$&${\color{red} 431}$&${\color{red}90.05}$&${\color{red}89.38}$\\
                              $ $&$1.5$&$0.9$&$ 369$&$90.03$&$90.13$&$ 400$&$90.03$&$89.17$&${\color{red} 349}$&${\color{red}90.00}$&${\color{red}90.56}$&${\color{red} 380}$&${\color{red}90.01}$&${\color{red}89.51}$\\
                              $ $&$ $&$1.2$&$ 376$&$90.02$&$90.24$&$ 408$&$90.04$&$89.28$&${\color{red} 351}$&${\color{red}90.07}$&${\color{red}90.32}$&${\color{red} 382}$&${\color{red}90.02}$&${\color{red}89.31}$\\
\hline
 \end{tabular} \caption{Estimated sample size at the nominal $90\%$ power and simulated power (SIM) at the calculated  size in designing a new GCD superiority trial\newline
[1] SIM is evaluated using $40,000$ simulated trials.\newline
[2] Losses to follow-up  are exponentially distributed with mean $1/\delta=4$ years (annual dropout rate $22.1\%$) in both arms.}\label{supres}
\end{center}
\end{table}

  \begin{table}[h]
\begin{center}
\begin{tabular}{l@{\extracolsep{5pt}}c@{\extracolsep{5pt}}c@{\extracolsep{5pt}}c@{\extracolsep{5pt}}c@{\extracolsep{5pt}}c@{\extracolsep{5pt}}c@{\extracolsep{5pt}}c@{\extracolsep{5pt}}c@{\extracolsep{5pt}}c@{\extracolsep{5pt}}c@{\extracolsep{5pt}}c@{\extracolsep{5pt}}c@{\extracolsep{5pt}}c@{\extracolsep{5pt}}c@{\extracolsep{5pt}}c@{\extracolsep{5pt}}c@{\extracolsep{5pt}}c@{\extracolsep{5pt}}c@{\extracolsep{5pt}}c@{\extracolsep{5pt}}c@{\extracolsep{5pt}}c@{\extracolsep{5pt}}c} \\\hline 
&&  \multicolumn{8}{c}{Unequal dispersion$^{(a)}$} &  \multicolumn{8}{c}{Unequal dropout$^{(b)}$}  \\ \cline{3-10}\cline{11-18}
&& && \multicolumn{3}{c}{design 1 }  & \multicolumn{3}{c}{design 2}  & && \multicolumn{3}{c}{design 1 } &  \multicolumn{3}{c}{design 2 } \\\cline{5-7}\cline{8-10}\cline{13-15}\cline{16-18}
  &&& & total &\multicolumn{2}{c}{power ($\%$)}    & total  & \multicolumn{2}{c}{power ($\%$)} &&& total & \multicolumn{2}{c}{power ($\%$)} & total  & \multicolumn{2}{c}{power ($\%$)}  \\\cline{6-7}\cline{9-10}\cline{14-15}\cline{17-18}
 $\psi$ & $\nu$ & $\kappa_0$ & $\kappa_1$ & size  & nominal & SIM    & size &  nominal & SIM &$\kappa_0$ & $\kappa_1$&  size  & nominal & SIM    &size  & nominal & SIM \\\hline

  1.1&0.9&0.4&0.8&$ 324$&$90.08$&$91.12$&${\color{red} 292}$&${\color{red}90.05}$&${\color{red}90.95}$&0.4&0.4&$ 287$&$90.06$&$90.89$&${\color{red} 254}$&${\color{red}90.01}$&${\color{red}90.57}$\\
                           &&0.4&1.2&$ 358$&$90.02$&$91.07$&${\color{red} 328}$&${\color{red}90.08}$&${\color{red}91.10}$&0.8&0.8&$ 356$&$90.02$&$90.55$&${\color{red} 326}$&${\color{red}90.07}$&${\color{red}90.65}$\\
                           &&0.8&1.2&$ 393$&$90.04$&$90.84$&${\color{red} 363}$&${\color{red}90.03}$&${\color{red}90.80}$&1.2&1.2&$ 426$&$90.06$&$90.46$&${\color{red} 397}$&${\color{red}90.04}$&${\color{red}90.26}$\\
                           1.1&1.2&0.4&0.8&$ 330$&$90.06$&$90.79$&${\color{red} 287}$&${\color{red}90.01}$&${\color{red}90.69}$&0.4&0.4&$ 292$&$90.05$&$90.37$&${\color{red} 248}$&${\color{red}90.01}$&${\color{red}90.36}$\\
                           &&0.4&1.2&$ 365$&$90.03$&$91.02$&${\color{red} 324}$&${\color{red}90.03}$&${\color{red}90.99}$&0.8&0.8&$ 363$&$90.06$&$90.58$&${\color{red} 322}$&${\color{red}90.03}$&${\color{red}90.66}$\\
                           &&0.8&1.2&$ 400$&$90.00$&$90.75$&${\color{red} 361}$&${\color{red}90.04}$&${\color{red}90.69}$&1.2&1.2&$ 434$&$90.06$&$90.29$&${\color{red} 396}$&${\color{red}90.04}$&${\color{red}90.33}$\\
                           1.5&0.9&0.4&0.8&$ 265$&$90.04$&$90.82$&${\color{red} 243}$&${\color{red}90.09}$&${\color{red}90.94}$&0.4&0.4&$ 229$&$90.06$&$90.55$&${\color{red} 206}$&${\color{red}90.11}$&${\color{red}90.59}$\\
                           &&0.4&1.2&$ 300$&$90.07$&$91.26$&${\color{red} 278}$&${\color{red}90.02}$&${\color{red}91.12}$&0.8&0.8&$ 298$&$90.01$&$90.36$&${\color{red} 277}$&${\color{red}90.05}$&${\color{red}90.47}$\\
                           &&0.8&1.2&$ 334$&$90.00$&$90.70$&${\color{red} 314}$&${\color{red}90.06}$&${\color{red}90.89}$&1.2&1.2&$ 368$&$90.06$&$90.33$&${\color{red} 348}$&${\color{red}90.02}$&${\color{red}90.26}$\\
                           1.5&1.2&0.4&0.8&$ 270$&$90.03$&$90.63$&${\color{red} 240}$&${\color{red}90.02}$&${\color{red}90.62}$&0.4&0.4&$ 233$&$90.05$&$90.51$&${\color{red} 202}$&${\color{red}90.08}$&${\color{red}90.07}$\\
                           &&0.4&1.2&$ 306$&$90.08$&$91.13$&${\color{red} 277}$&${\color{red}90.04}$&${\color{red}90.72}$&0.8&0.8&$ 304$&$90.05$&$90.63$&${\color{red} 276}$&${\color{red}90.08}$&${\color{red}90.48}$\\
                           &&0.8&1.2&$ 341$&$90.05$&$90.67$&${\color{red} 314}$&${\color{red}90.06}$&${\color{red}90.79}$&1.2&1.2&$ 375$&$90.06$&$90.37$&${\color{red} 349}$&${\color{red}90.01}$&${\color{red}90.45}$\\

\hline

 \end{tabular} \caption{Estimated sample size  and simulated power (SIM) at the nominal $90\%$ power in the presence of unequal dropout or dispersion \newline
{\normalsize
[1] SIM is evaluated using $10,000$ simulated trials\newline
[2] The treatment allocation ratio is $1:1$\newline
 $^{(a)}$   Losses to follow-up  are exponentially distributed with mean $1/\delta=4$ years (annual dropout rate $22.1\%$) in both arms.\newline
 $^{(b)}$ Losses to follow-up  are exponentially distributed  with $\delta_1=0.15$  and $\delta_0=0.35$ (annual dropout rates $13.9\%$ and $29.5\%$)  in the two arms. \newline
[3] {\color{red} The sample size and nominal power estimates are updated for design 2 with staggered entry. The simulated power  may be different from the previously reported values after re-runing the simulation for design 1.}} 
 }\label{supres2}
\end{center}
\end{table}

\subsection{Example 2}
We conduct simulations to assess the performance of the proposed method in the presence of unequal dispersion or differential  dropout.
Two scenarios are considered. In one scenario, the dispersion parameters in the two groups are different.
In the other scenario, we assume different  loss to follow-up distributions for the two groups.
The setup is otherwise similar to that in the example $1$. 
The parameter values and simulation results are presented in Table \ref{supres2}. The performance of the power and sample size  method is almost as good as that in Example $1$.

\subsection{Example 3}
Simulation is conducted to assess the proposed sample size method for NI and equivalence trials. For illustration purposes, we assume a piecewise constant event rate function for the control arm
$\lambda_0(t)=1.0 I(0\leq t <0.4) + 1.25 I(0.4 \leq t <0.8) + 1.5 I (0.8\leq t\leq 1)$,  the event rate ratio between the active and control arm is $\exp(\beta)=\lambda_1(t)/\lambda_0(t)=0.9$ or $1.0$, and the dispersion parameter is
$\kappa=0.8$ or $1.2$. Only design $1$ is considered, and the planned treatment duration is $\tau_c=1$ year for all patients. The treatment allocation ratio is $1:1$.
The loss to follow-up is  exponentially distributed with mean $1/\delta=4$ years. The margin is $M_0=1.25$ in the NI trials, and  $(M_l,M_u)=(0.75,1.25)$ in the equivalence trials.

Table \ref{nires} reports the sample size and power estimates at the target $80\%$ power and one-sided type I error $\alpha/2=0.025$.   
The empirical  power is evaluated based on $10,000$ simulated trials. 
 There is more than $95\%$ chance
that the simulated power lies within $2 \sqrt{ 0.8*0.2/10000}=0.8\%$ of the true power.
The simulated power at the calculated sample size is generally close to the target $80\%$ power, indicating the accuracy of the proposed method.
 
 \begin{table}[h]
\begin{center}
\begin{tabular}{l@{\extracolsep{5pt}}c@{\extracolsep{5pt}}c@{\extracolsep{5pt}}c@{\extracolsep{5pt}}c@{\extracolsep{5pt}}c@{\extracolsep{5pt}}c@{\extracolsep{5pt}}c@{\extracolsep{5pt}}c@{\extracolsep{5pt}}c@{\extracolsep{5pt}}c@{\extracolsep{5pt}}c@{\extracolsep{5pt}}c@{\extracolsep{5pt}}ccccc} \\\hline 
\multicolumn{5}{c}{NI trials $^{(a)}$}  &&  \multicolumn{5}{c}{Equivalence trials $^{(b)}$ } \\\cline{1-5}\cline{7-11}
  &&total &\multicolumn{2}{c}{power ($\%$)}    &&&&total &\multicolumn{2}{c}{power ($\%$)}  \\\cline{4-5}\cline{10-11}
 $\kappa$ & $\exp(\beta)$&   size & nominal & SIM   &&  $\kappa$ & $\exp(\beta)$&  size & nominal & SIM \\\hline   
         $0.8$&$ 0.9$&$ 547$&$80.00$&$80.26$&&  $0.8$&$0.9$&$1781$&$80.02$&$79.49$\\
     & $ 1.0$&$1153$&$80.03$&$79.96$&&   &$1.0$&$1262$&$80.02$&$79.58$\\     
           $1.2$&$ 0.9$&$ 675$&$80.04$&$80.32$&&  $1.2$&$0.9$&$2195$&$80.01$&$81.20$\\   
 & $ 1.0$&$1429$&$80.02$&$80.31$&&    &$1.0$&$1564$&$80.01$&$80.47$\\ 
\hline
 \end{tabular} \caption{ Estimated sample size at the nominal $80\%$ power and simulated power (SIM) at the calculated sample size based on $10,000$ NI or 
equivalence trials  \newline
$^{(a)}$ NI margin is $M_0=1.25$ \newline
$^{(b)}$  Equivalence margin is $(M_l,M_u)=(0.75,1.25)$
}\label{nires}
\end{center}
\end{table}

\section{Discussion}
We derive the power and sample size formulae for comparing recurrent rates  in superiority, NI and equivalence trials using the robust Wald test from the AG model. 
The method allows the dispersion parameter, dropout rate, and/or sample size to differ between treatment groups.
Numerical examples demonstrate the accuracy of the proposed method in moderate-to-large samples. 
It is always recommended to run  simulation studies to  verify the power  calculation particularly when the sample size  is relatively small. 

We calculate the variance $V_\beta$ and  the sample size at given event rate function, dispersion parameter and dropout rate. These parameters may be estimated 
from the historical trials using parametric methods.  It is flexible to adjust the parameter values and conduct sensitivity analyses to examine
how the sample size estimates vary with these parameter values.
 Please see Example $1$ for illustration.
It is  possible to estimate $V_\beta$ from the historical trials by  nonparametric methods \cite{song:2008}. However, the nonparametric approach  may require that 
the new trial is sufficiently similar to the historical trial in terms of the study population, treatment duration, drop rates, etc.

The robust  AG approach has  several limitations. 
First, the AG model 
 uses a common baseline hazard function for all events, and assumes  that the risk of an event is unaffected by any early events that occur within the same subject. Therefore, the AG model
is not suitable if the  occurrence of  early events increases the risk for subsequent ones  \cite{wang:2009, amorim:2015}.
The AG model provides a convenient way to  estimate an overall treatment effect, but it would be difficult to estimate  the event specific treatment effect \cite{wang:2009, amorim:2015}, which is useful for studying 
whether  the treatment effect reduces after the patients experience one or more events.
Second, when the sample size is small, the sandwich variance estimator tends to underestimate  the true variance and have large sampling variability, leading to inflated type I error rate \cite{kauermann:2001, lu:2007}. 
In the GEE methodology,  the bias corrected sandwich variance estimator has been proposed for small sample inferences \cite{kauermann:2001, lu:2007, guo:2005}.
It is possible to extend the bias correction method to the analysis of recurrent events. An alternative strategy for the analysis of small trials is to use the robust score test instead of the robust Wald test \cite{guo:2005}.

\appendix
\section{Appendix: Technical details}
\subsection{A brief proof of equations \eqref{sigma1} and \eqref{sigma2}}\label{proofsig}
By Lin {et al}  \cite{lin:2000}, $\hat{V}_\beta$ is a consistent estimate of $V_\beta$
\begin{equation}\label{sigma10}
V_\beta = \frac{\text{E}[\text{E}(U_i^2|x_i)]}{\text{E}^2(I_\beta)}=\frac{ p_1\Sigma_1+p_0\Sigma_0}{\left( \int_{0}^\tau \omega_1(t)\omega_0(t)[p_1\pi_1(t)d\Lambda_1+p_0\pi_0(t)d\Lambda_0] \right)^2},
\end{equation}
where  $d\, M_i(t)=dN_i(t)-Y_i(t) \exp(\beta x_i) d\Lambda_0(t)$, $U_i=\int_{0}^\tau [x_i -\bar{x}(\beta,t)] d\,{M}_i(t)$, $\Sigma_g=\text{E}(U_i^2|x_i=g)$,   and $ I_\beta = n^{-1} \sum_{i=1}^{n} \int_0^\tau [\bar{x}(\beta,t)-\bar{x}^2(\beta,t)]dN_{i}(t)$.
By Lemma $1$ in the web-based 
supplementary material of Song {\it et al} \cite{song:2008}, we get
\begin{equation}\label{sigma11}
\Sigma_g= \text{E}\left[\int_{0}^\tau (g-\bar{x}(\beta, t))dM_i(t)\int_{0}^\tau (g-\bar{x}(\beta, s))dM_i(s)\right]=A_g+\kappa B_g
\end{equation}
 for subjects in group $g$, where
\begin{eqnarray*}\label{meanfun}
\begin{aligned}
A_g& =\text{E}\left[ \int_{0}^{\tau}Y_i(t) \omega_{g}^2(t) d\Lambda_{g}(t) \right] = \int \omega_{g}^2(t) \pi_{g}(s)d\Lambda_{g}(t), \\
B_g& = \text{E}\left[ \int_{0}^\tau \int_{0}^\tau Y_i(t)Y_i(s) \omega_{g}(t)\omega_{g}(s) d\Lambda_{g}(t)d\Lambda_{g}(s)\right ] =2 \int_{s=0}^\infty\left[\int_{t=0}^s \omega_{g}(t) d\Lambda(t)\right] \pi_g(s) \omega_{g}(s)d\Lambda_g(s).
\end{aligned}
\end{eqnarray*}
 Inserting Equation \eqref{sigma11} into Equation \eqref{sigma10} yields Equation \eqref{sigma1}.

 Equation \eqref{sigma2} holds under equal dropout since  $\omega_0(t)\equiv p_1\exp(\beta)/D$, $\omega_1(t)\equiv p_0/D$,
$ I_\beta  =p_0\omega_0(t)E_0=p_1\omega_1(t)E_1$, $A_i=\omega_{i}^2(t) E_i$,  $B_i=2 \omega_{i}^2(t) F_i$, 
$E_1=E_0\exp(\beta)$, $F_1=F_0\exp(2\beta)$ and $F_1/E_1^2=F_0/E_0^2$,
where $D=p_0+p_1\exp(\beta)$.

\subsection{Asymptotic variance expressions in two designs under equal dropout}\label{analytvar}

\subsubsection{Design 1}\label{design1}
The  planned treatment duration is $\tau_c$ years for each subject (the accrual period is irrelevant in the sample size calculation). The loss to follow-up is exponentially distributed with mean $\delta^{-1}$.
The probability that a subject is in the trial at time $t$ after randomization is
 $\pi(t)= \exp(-\delta t)$.

{\flushleft{\bf Weibull event rate function}}\\
Suppose the  rate function is $\lambda_0(t)=\psi \nu t^{\nu-1}$ and the mean function is $\Lambda_0(t)=\psi t^{\nu}$ for the recurrent event in the control group, where $\psi$ is a scale parameter and $\nu$ is a shape parameter. 
Let  $\text{IG}(\nu,a) =\int_{t=0}^a t^{\nu-1}\exp(-t)dt$ be the incomplete gamma function and $\text{IG}(\nu,a,b)=\int_{t=b}^a t^{\nu-1}\exp(-t)dt$. We have
$$E_0^{\text{I}}=\int_{0}^{\tau_c} \pi(t) d\Lambda_0(t)=\psi \nu \int_{0}^{\tau_c} \exp(-\delta t)t^{\nu-1}dt=\begin{cases} \frac{\psi \nu}{\delta^\nu}\text{IG}(\nu,\delta\tau_c) & \text{ if } \delta\neq 0\\
\psi \tau_c^\nu  & \text{ at } \delta=0,
\end{cases}$$
$$F_0^{\text{I}}=\int_{0}^{\tau_c} \pi(t) \Lambda_0(t) d\Lambda_0(t)=\psi^2 \nu \int_{0}^{\tau_c} \exp(-\delta t)t^{2\nu-1}dt= \begin{cases} \frac{\psi^2 \nu}{\delta^{2\nu}} \text{IG}(2\nu,\delta\tau_c) & \text{ if } \delta\neq 0\\
                \psi^2 \tau_c^{2\nu}/2     & \text{ at } \delta=0.
\end{cases}  $$

{\flushleft{\bf Piecewise constant event rate function}}\\
Let $\lambda_0(t)=\sum_{k=1}^d \tilde\lambda_k I(l_{k-1}\leq t<l_k)$, where $l_0=0$, $l_d=\tau_c$. Then $\Lambda_0(t)=\Lambda_0(l_{k-1})+\tilde\lambda_k (t-l_{k-1})$ when $l_{k-1}\leq t<l_k$.
Let $\Delta_k=l_k-l_{k-1}$ and $G_{km}=\int_{l_{k-1}}^{l_k} \exp[-\delta (t-l_{k-1})] (t-l_{k-1})^m dt= \int_{0}^{\Delta_k} \exp[-\delta t] t^m dt$ for $m=0,1,2$. Then
\begin{equation}\label{gfunc}
\begin{cases} 
G_{k0}=\frac{1-\exp(-\delta \Delta_k)}{\delta},\,\, G_{k1}= \frac{1-(1+\delta \Delta_k) \exp(-\delta \Delta_k)}{\delta^2},\,\,G_{k2}=\frac{2-(\delta^2\Delta_k^2+2\delta\Delta_k+2)\exp(-\delta\Delta_k)}{\delta^3} & \text{ if } \delta >0 \\
G_{k0}= \Delta_k,\,\, G_{k1}= \frac{\Delta_k^2}{2},\,\,G_{k2}=\frac{\Delta_k^3}{3} & \text{ at } \delta =0 
\end{cases}
\end{equation} 

We have
$$E_0^{\text{I}}=\int_{0}^{\tau_c} \pi(t) d\Lambda_0(t)= \sum_{k=1}^d \tilde\lambda_k \int_{l_{k-1}}^{l_k} \pi(t) dt= \begin{cases}  \sum_{k=1}^d \tilde{\lambda}_k \exp[-\delta l_{k-1}] G_{k0}  & \text{ if } \delta\neq 0\\ 
\Lambda(\tau_c)  & \text{ at } \delta=0,
\end{cases}  $$
$$F_0^{\text{I}}=\int_{0}^{\tau_c} \pi(t)\Lambda_0(t) d\Lambda_0(t)= \sum_{k=1}^d \tilde\lambda_k \int_{l_{k-1}}^{l_k}  \pi(t) \Lambda_0(t) dt= \begin{cases}  \sum_{k=1}^d \tilde{\lambda}_k \exp[-\delta l_{k-1}] [\Lambda(l_{k-1}) G_{k0}+ \tilde\lambda_k G_{k1}]   & \text{ if } \delta\neq 0\\ 
\Lambda^2(\tau_c)/2  & \text{ at } \delta=0.
\end{cases}$$

\subsubsection{Design 2}\label{design2}
Subjects are enrolled during  an accrual period of $\tau_a$ years, and followed for an additional $\tau_c$ years after the closure of recruitment. The total study duration is $\tau=\tau_a+\tau_c$ years.
 Suppose the entry time for a subject is distributed with density function given by
  $$f(e_{i}) =\frac{ \eta \exp(-\eta e_{i})}{1-\exp(-\eta\tau_a)}, \text{ where } 0\leq e_{i}\leq \tau_a.$$
The entry distribution is convex
(faster patient entry at the beginning) if $\eta > 0$, and concave (lagging patient entry) if $\eta < 0$, and uniform $f(e_{i})=1/\tau_a$ if $\eta\rightarrow 0$.
In terms of  the sample size calculation, design $1$ can be viewed as a special case of design $2$ by setting $\tau_a=0$.

Given the entry time $e_{i}$, the maximum follow-up for an individual is $\tau-e_{i}$. We assume the loss to follow-up is exponentially distributed  with mean $\delta^{-1}$.
The probability that a subject is still in the trial at time $t$ after randomization is 
\begin{eqnarray}\label{surv2}
\begin{aligned}
\pi(t) & =\Pr(T_{i} >t)=\Pr(T_{i} >t | e_{i}+t\leq \tau) \Pr(e_{i}+t\leq \tau) +\Pr(T_{i} >t | e_{i}+t > \tau ) \Pr(e_{i}+t> \tau)  \\
& = \begin{cases} \exp(-\delta t) \text{ if } t\leq \tau_c \\
  \exp(-\delta t) {\color{red} \frac{1-\exp[-\eta(\tau-t)]}{1-\exp(-\eta\tau_a)}}   \text{ if } \tau_c < t\leq \tau.
\end{cases}
\end{aligned}
\end{eqnarray}
When $\eta\rightarrow 0$, we shall replace ${\color{red} \frac{1-\exp[-\eta(\tau-t)]}{1-\exp(-\eta\tau_a)}}$ by its limiting value ${\color{red} (\tau-t)/\tau_a}$ in Equation \eqref{surv2}. 
In design 2, it is easy to see that 
$$ \int_{0}^\tau \pi(t) d\Lambda_0(t) =E^{\text{I}} +\int_{\tau_c}^\tau \pi(t) d\Lambda_0(t) \text{ and }  \int_{0}^\tau \pi(t) \Lambda_0(t) d\Lambda_0(t) =F_0^{\text{I}} +\int_{\tau_c}^\tau \pi(t) \Lambda_0(t) d\Lambda_0(t), $$
where $E_0^{\text{I}}$ and $F_0^{\text{I}}$ are defined in Appendix \ref{design1}. Below we give analytic expression for $\int_{\tau_c}^\tau \pi(t) d\Lambda_0(t) $ and 
$\int_{\tau_c}^\tau \pi(t) \Lambda_0(t) d\Lambda_0(t)$ at $\eta=0$. The expressions are omitted when $\eta\neq 0$ due to limited space.

{\flushleft{\bf Weibull event rate function}}\\
Suppose $\lambda_0(t)=\psi \nu t^{\nu-1}$. When  $\eta= 0$, we get 
\begin{eqnarray}
\begin{aligned}
\int_{\tau_c}^\tau \pi(t) d\Lambda_0(t) &= \begin{cases} {\color{red} \frac{\tau\psi \nu}{\tau_a \delta^{\nu}} \text{IG}(\nu,\delta\tau,\delta\tau_c) - \frac{\psi \nu}{\tau_a \delta^{\nu+1}} \text{IG}(\nu+1,\delta\tau,\delta\tau_c) } & \text{ if } \delta>0  \\
                             {\color{red}       \frac{\tau\psi }{\tau_a} [\tau^{\nu}-\tau_c^{\nu}]   -   \frac{\psi \nu}{\tau_a(\nu+1)} [\tau^{\nu+1}-\tau_c^{\nu+1}] }&  \text{ at } \delta=0      \end{cases} \\
\int_{\tau_c}^\tau \pi(t)\Lambda_0(t) d\Lambda_0(t)&= 
        \begin{cases} { \color{red}     \frac{\tau\psi^2 \nu}{\tau_a \delta^{2\nu}} \text{IG}(2\nu,\delta\tau,\delta\tau_c) - \frac{\psi^2 \nu}{\tau_a \delta^{2\nu+1}} \text{IG}(2\nu+1,\delta\tau,\delta\tau_c) } & \text{ if } \delta>0  \\
                     { \color{red}            \frac{\tau\psi^2 }{2\tau_a} [\tau^{2\nu}-\tau_c^{2\nu}] -     \frac{\psi^2 \nu}{\tau_a (2\nu+1)} [\tau^{2\nu+1}-\tau_c^{2\nu+1}]}  & \text{ at } \delta=0      \end{cases} \\
\end{aligned}
\end{eqnarray}

{\flushleft{\bf Piecewise constant event rate function}}\\
Suppose $\lambda_0(t)= \sum_{k=1}^{d^*} \tilde\lambda_k I(l_{k-1}\leq t < l_k)$, where $l_{d^*}=\tau=\tau_a+\tau_c$ and $l_{d}=\tau_c$. 
For notational convenience, if $\tau_c$ is not a knot, it can be added as a knot. 
When $\eta=0$,
\begin{eqnarray*}
\begin{aligned}
&\int_{\tau_c}^\tau \pi(t) d\Lambda_0(t)={\color{red} \sum_{k=d+1}^{d^*}\int_{l_{k-1}}^{l_k} \lambda_k \exp(-\delta t) \frac{\tau-t}{\tau_a} dt = \sum_{k=d+1}^{d^*}  \frac{ \lambda_k}{\tau_a} \exp(-\delta l_{k-1}) 
  \left[(\tau-l_{k-1}) G_{k0} - G_{k1} \right]}, \\
&\int_{\tau_c}^\tau \pi(t)\Lambda_0(t) d\Lambda_0(t)={\color{red} \sum_{k=d+1}^{d^*} \frac{\lambda_k}{\tau_a} \exp(-\delta l_{k-1}) \left\{ \Lambda_{k-1}[(\tau-l_{k-1})G_{k0}-G_{k1}] +\lambda_k [(\tau-l_{k-1})G_{k1}- G_{k2}] \right\}},
\end{aligned}
\end{eqnarray*}
where  $G_{k0}$,  $G_{k1}$ and  $G_{k2}$ are defined in Equation \eqref{gfunc}.
\bibliography{agsizenon} 
\end{document}